\documentclass[elsarticle,5p,showpacs,preprintnumbers,amsmath,amssymb]{elsarticle}

\usepackage{graphicx}
\usepackage{dcolumn}
\usepackage{bm}

\journal{Physica C}

\begin{document}

\title{Electronic Structure of Fe-Based Superconductors}

\author[djs]{D.J. Singh}

\address[djs]{Materials Science and Technology Division,
Oak Ridge National Laboratory, Oak Ridge, Tennessee 37831-6114}

\begin{abstract}
The electronic structure of the Fe-based superconductors is discussed,
mainly from the point of view of first principles calculations in relation
to experimental data. Comparisons and contrasts with cuprates are made.
The problem of reconciling experiments indicating an $s$ symmetry
gap with experiments indicating line nodes is discussed and a possible
resolution is given.
\end{abstract}


\maketitle

\section{Introduction}

Superconductivity is fundamentally an electronic phenomena,
and is in particular an instability of the metallic Fermi surface
due to interactions. Therefore understanding the electronic structure,
especially the low energy electronic structure at and near the Fermi 
energy, as well as the interactions that may lead to superconductivity
is essential to unraveling the physics of a superconducting material.
This includes developing understanding of the details of the interactions
in relation to competing or cooperating states such as magnetic orderings
and lattice distortions. The purpose of this paper is to discuss the
electronic structure of the Fe-based superconductors (FeSC),
\cite{kamihara-p,kamihara-a}
mainly from the
perspective of first principles calculations in relation to experiment.

\section{Crystal Structure and Chemistry}

Since the discovery of superconductivity in
electron doped oxy-pnictides, prototype
LaFeAs(O,F), high temperature superconductivity has been discovered in
three additional families of iron compounds:
ThCr$_2$Si$_2$ structure materials, prototype BaFe$_2$As$_2$, with
either hole or electron doping,
\cite{rotter-sc}
the LiFeAs family,
\cite{wang-lifeas,pitcher,tapp}
and the $\alpha$-PbO structure iron chalcogenides, prototype Fe$_{1+x}$Se.
\cite{hsu,mizuguchi}
Remarkably, superconductivity can be produced by
electron doping on the Fe site
itself, with either Co or Ni. 
\cite{sefat-co,sefat-co1,matsuishi}
In addition, superconductivity has been found in the corresponding pure
Ni based compounds,
both the oxy-pnictides and in the ThCr$_2$Si$_2$ structure.
\cite{watanabe-lanipo,ronning-ni,li-ni,watanabe2}
However, the superconductivity in these compounds can be understood as
ordinary electron-phonon superconductivity,
\cite{subedi-lanipo,subedi-banias,kurita-banias}
in contrast to the Fe-based materials, which cannot be understood
in this way.
\cite{boeri,mazin-spm}
The related Co compounds are not superconducting
and are either ferromagnetic or near
ferromagnetism. \cite{yanagi,sefat-baco}
Here we focus on the Fe-based phases.

\begin{figure}
\includegraphics[width=0.95\columnwidth]{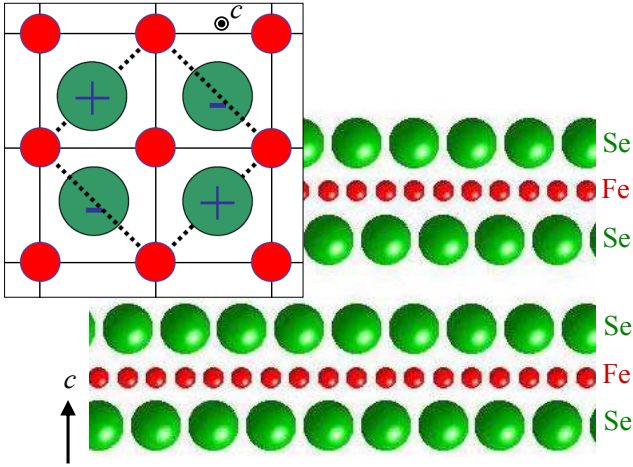}
\caption{(Color online)
Crystal structure of $\alpha$-FeSe shown along a (120) direction
using sphere sizes proportional to the Shannon ionic radii of
high spin Fe$^{2+}$ and Se$^{2-}$. The inset
(not to scale) shows the Fe plane with
coordinating Se atoms above (+) and below (-) the plane.
Note the resulting c(2x2) doubling of the cell (dashed line)
with respect to the Fe square lattice.}
\label{structure}
\end{figure}

The basic structural feature connecting these compounds is
the presence of square planar sheets of Fe coordinated tetrahedrally
by pnictogens or chalcogens and nominal Fe valence near Fe$^{2+}$.
We begin with FeSe, which is the simplest of the compounds. Its
structure is depicted in Fig. \ref{structure}.
This structure consists of square planes of Fe with Se atoms arranged
above and below the planes in such as way as to tetrahedrally coordinate
the iron.
The arrangement of Se above and below the Fe plane leads to a
$c$(2x2) doubling of the
unit cell compared to the Fe square lattice, so the actual unit
cell contains two Fe atoms.
The lattice may also be regarded as
a tetragonally distorted close packed lattice of Se, with Fe inside tetrahedral
holes arranged so that the Fe atoms are in a square plane.

The other FeSC compounds may be regarded
as based on the same square planar sheets with Se replaced by As
and counter-ions inserted in such a way as to maintain the nominal Fe valence.
For example, LiFeAs may be regarded as FeSe with Se replaced by As and
Li inserted between the layers.
In the ThCr$_2$Si$_2$ structure, e.g. BaFe$_2$As$_2$, the
FeAs planes have Ba between them, and the stacking is changed from simple
tetragonal, as in LiFeAs, FeSe and LaFeAsO, to body centered tetragonal,
thus providing a better coordination for Ba.

This type of structure
is an important difference from cuprates. In particular,
because of the larger size of Se and As anions, relative to O,
tetrahedral coordination is preferred, leading to a structure
composed of edge sharing tetrahedra. In cuprates the structural
motif is that of corner sharing octahedra. This is important for two
reasons. First of all, in materials built from corner sharing octahedra
the metal - metal distance is much longer than the metal ligand
distance (by a factor of two for an undistorted perovskite)
so that direct metal - metal interactions typically play a very minor
role compared to hopping through the ligands, e.g. in band formation
and magnetic interactions. Secondly, structures built from corner
sharing octahedra almost invariably are prone to structural distortions e.g.
in order to accommodate counter ions with sizes that are not perfectly
matched to their sites.
Thus the ideal cubic perovskite
($Pm3m$, BaZrO$_3$) structure is relatively uncommon, while distorted
perovskite structures (e.g. $Pnma$, CaTiO$_3$) are very common.
These distortions typically couple strongly to electronic structure,
and to magnetic and other properties, as might be anticipated based on the
fact that metal - O - metal hopping plays a key role in band formation.
Similarly, some of the most studied cuprates, particularly the
Bi compounds (e.g. Bi$_2$Sr$_2$CaCu$_2$O$_8$, BISCO) and the ``214" family
(e.g. La$_{2-x}$Sr$_x$CuO$_4$) show strong lattice distortions that
greatly complicate the interpretation of experimental results. One
instructive example is provided by the long-standing misinterpretation
of the shadow bands in BISCO as being a novel manifestation of strong
correlation effects, when in fact they are just a band structure
effect arising from the complex lattice distortion in that compound.
\cite{singh-bi,koitzsch-bi,nakayama-bi}
The key point is that the structures of the FeSC are much simpler,
and in particular these compounds are not prone to complex lattice
distortions. This should greatly facilitate experimental studies of
these materials and comparison of theoretical and experimental results.
Also, we note that the structure places the Fe atoms closer together
than in a perovskite, so that direct Fe - Fe interactions may be
(and are, see below) important.

\begin{figure}
\includegraphics[height=0.95\columnwidth,angle=270]{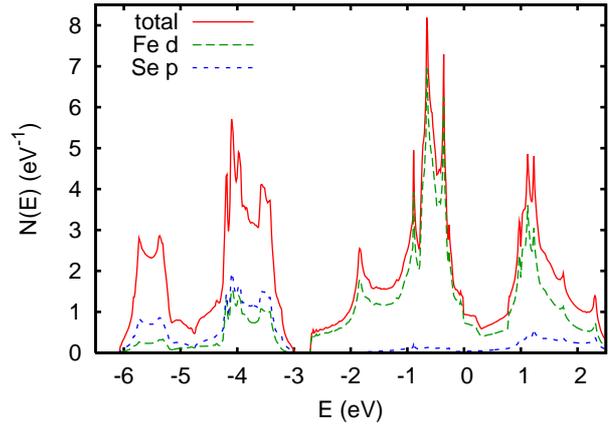}
\caption{(Color online)
Calculated electronic density of states for FeSe and projections
onto atom centered spheres of radius 2.1 Bohr
following Ref. \cite{subedi-fese}. Note that the
$p$ orbitals of Se are extended, and so the projection is proportional
to, but underestimates the Se $p$ contribution.
}
\label{dos-fese}
\end{figure}

\begin{figure}
\includegraphics[height=0.95\columnwidth,angle=270]{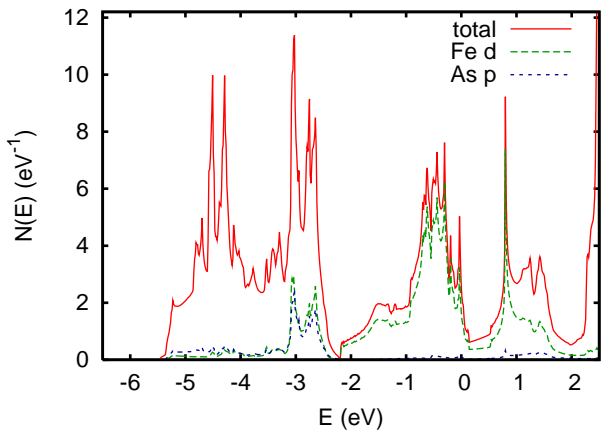}
\caption{(Color online)
Calculated electronic density of states for LaFeAsO and projections
onto atom centered spheres of radius 2.1 Bohr as in Fig. \ref{dos-fese},
following Ref. \cite{singh-du}.
}
\label{dos-lfa}
\end{figure}

The electronic structures of the various compounds as obtained
within density functional theory have been reported by a number
of authors.
\cite{mazin-spm,lebegue,singh-du,ishibashi,
yildirim,yin,mazin-mag,che,kamihara-el,lu-pe,
takenaka,singh-bla,zhang-el,nekrasov-el,ma-mag, nekrasov-lfa,subedi-fese,
xu-fm,dong-sdw,vildosola,jeevan,malaeb,nekrasov-comp,shein-comp,nakamura,
belashchencko,sushko}
While these differ in detail a number of common features
are present.
Fig. \ref{dos-fese} shows the electronic
density of states (DOS) of FeSe
as obtained in the local density approximation (LDA) without including
magnetism.
\cite{subedi-fese}
As may be seen, the main Se $p$ bands are well below the Fermi energy
(between 3 eV and 6 eV binding energy) and
the DOS from -2.5 eV to 2 eV is dominated by Fe $d$ states,
with only a modest admixture of Se character, similar to the level
of covalency in typical oxides. The arsenides are similar as shown
in Fig. \ref{dos-lfa} for LaFeAsO. In that compound the main As $p$
bands occur between $\sim$ -5.5 and -2 eV with respect to the Fermi
energy, $E_F$ and again the bands near $E_F$ are dominated by Fe $d$
character. 
This implies that from a crystal chemical point of view the Se and
As are anionic, Se$^{2-}$ and As$^{3-}$ and that the electronic
structure near $E_F$ should be regarded as derived from
metallic square lattice sheets of Fe$^{2+}$ embedded in tetrahedral
holes of the anion lattice.
This picture has been confirmed by core-level and valence photoemission
experiments, which show the bands near $E_F$ to be predominantly Fe
in character. \cite{koitzsch-fe}
We note that this modest $d$ - $p$ hybridization is common to the
Fe, Co and Ni materials in this structure and is in strong contrast
with the behavior of the corresponding Mn compounds. \cite{an}

An examination of the DOS (Figs. \ref{dos-fese} and \ref{dos-lfa})
shows that Fe $d$ manifold is split into to main peaks. These
are separated by a rather prominent pseudogap, with $E_F$ occurring
towards the bottom.
This pseudogap occurs
at an electron count of six per Fe, corresponding to the $d$ electron
count of Fe$^{2+}$. Importantly, a tetrahedral crystal field scheme,
such as might arise if the Fe - As (Se) interactions were dominant,
would have a gap at four electrons per Fe, since in a tetrahedral
ligand field the $e_g$ levels would be below the $t_{2g}$ levels.
Instead, the position of the pseudogap shows the importance of direct Fe-Fe
interactions in the formation of the band structure.
This in contrast to cuprate superconductors, where the Cu ions are
contained within distorted corner sharing octahedra, and hopping through
O plays the critical role in band formation, conduction and magnetic
exchange interactions.

Calculations have also been performed using methods that incorporate
an explicit Hubbard-like Coulomb term, $U$, via the dynamical mean field
theory (DMFT). \cite{haule,craco}
Those calculations, as is usual when incorporating an additional on-site
Coulomb repulsion, yield a strong shift of Fe $d$ spectral weight
away from the region around $E_F$ to Hubbard bands at binding energies
corresponding to $U$. Additionally, it is claimed that the metallic
state is destroyed in favor of a total incoherent state. \cite{craco}
This disagrees qualitatively with experimental data, which show a
metallic state, including observation of quantum oscillations,
\cite{sebastian,coldea,sugawara}
metallic-like band dispersions around $E_F$,
\cite{lu-pe,liu-pe}
and critically the non-observation of Hubbard bands. \cite{kurmaev}
From this perspective the FeSC behave very
differently from cuprate superconductors.
At first sight, this may seem surprising as one might wonder how adding
the $U$ could degrade results. However, it is to be emphasized that
density functional theory does already contain some correlations
and as a result the addition of $U$ can lead to double counting of
correlations with detrimental effects in band-like metals.
It should also be noted that arguments as to why correlation effects
should be weak have been advanced by Nakamura and co-workers, \cite{nakamura-2}
and also by
Anisimov and co-workers, \cite{anisimov}
based on a Wannier function analysis.
In any case, considering the non-observation of the Hubbard bands
and the non-observation of the associated strong
spectral weight shifts into them as predicted by current DMFT
calculations with substantial $U$ we do not discuss these further here.

\begin{figure}
\includegraphics[height=0.95\columnwidth,angle=270]{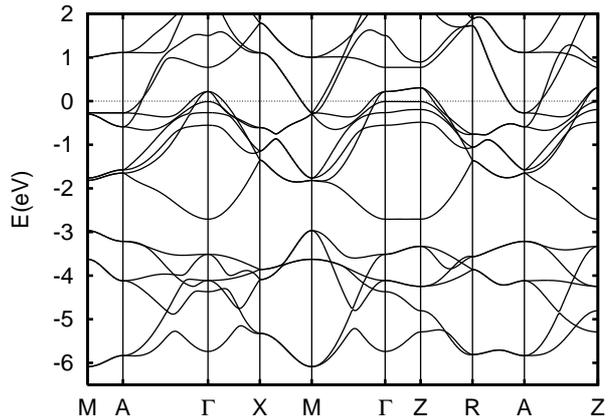}
\caption{
Calculated band structure for FeSe
following Ref. \cite{subedi-fese}.
}
\label{band-fese}
\end{figure}

\section{Band Structure and Fermi Surface}

\begin{figure}
\includegraphics[width=1.00\columnwidth]{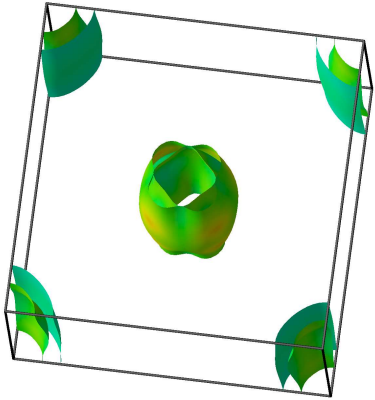}
\caption{(Color online)
Calculated Fermi surface for FeSe
following Ref. \cite{subedi-fese}.
}
\label{fermi-fese}
\end{figure}

The non-spin-polarized
LDA band structure for FeSe
is shown in Fig. \ref{band-fese},
following Ref. \cite{subedi-fese}.
The corresponding Fermi surface is shown in Fig. \ref{fermi-fese}.
As may be seen the band structure is semi-metallic, with small
Fermi surface sections, in contrast to cuprates. \cite{pickett-htc}
While the layered crystal structure is reflected in the cylindrical
shape of the Fermi surface, there is clear three dimensionality.
The amount of three dimensionality varies from compound to compound,
but is particularly pronounced in ThCr$_2$Si$_2$
structure BaFe$_2$As$_2$,
\cite{singh-bla,nekrasov-comp}
where one of the hole sections shows rather strong k$_z$ dispersion 
near $k_z$=1/2.
Nonetheless, considering the Fermi surface, the FeSC
are substantially less anisotropic than the cuprates,
including the simplest tetragonal one layer compounds, Tl$_2$Ba$_2$CuO$_6$,
\cite{singh-tl} and HgBa$_2$CuO$_4$, \cite{singh-hg}
and even considering YBa$_2$Cu$_3$O$_7$,
which is one of the least anisotropic cuprates.

The relatively low anisotropy of the FeSC is of potential practical
importance. In particular, flux pinning is an important issue
in applications of superconductivity, and this is greatly facilitated
in low anisotropy materials.
The lower anisotropy of FeSC is supported by several experiments,
and in fact favorable results for flux pinning have been obtained
in FeSC samples.
\cite{jia-anis,wang-anis,altarawneh,yamamoto}

The basic structure of the Fermi surface is similar in all the FeSC.
It consists of two electron cylinders centered at the zone corner,
compensated by hole sections around the zone center. Both are primarily
derived from Fe $d_{xz}$ and $d_{yz}$ orbitals (with the reference frame
have $z$ normal to the Fe planes). In addition, in some compounds,
an additional heavy $d_{z^2}$ derived heavy hole section is predicted
at the zone center. \cite{lebegue,singh-du}
As mentioned, there is hybridization between Fe and the ligands, albeit
modest. In particular, the $d$ bands at $E_F$ have antibonding character
with the As / Se ligands.

The electron sections may be regarded \cite{mazin-spm,kuroki-spm}
as deriving from the 2D
zone boundary ($\pi$,0) and (0,$\pi$) $d_{xz}$ and $d_{yz}$ bands in the
simple one Fe unit cell that would exist without the As / Se atoms
above and below the plane.
Both of these points are folded to the 2D
zone corner ($\pi$,$\pi$) point in the $c$(2x2) unit cell of the compound.
Since these bands arise from the zone boundary of the Fe lattice, they
have antibonding character between the Fe atoms and are more dispersive
than the $d_{xz}$/$d_{yz}$ bands at the zone center, which are
bonding in character and heavier.

This basic Fermi surface structure with small disconnected
hole and electron Fermi surfaces at the zone center and
zone corner, respectively, is confirmed by photoemission
experiments.
\cite{lu-pe,liu-pe,ding-pe,zhao-pe,kondo-pe,wray-pe}
Furthermore, high field measurements show two gap superconductivity,
\cite{hunte}
with the implication that both sets of Fermi surface are present
and play a role in the superconducting phase.

Finally, it is important to emphasize that although the Fermi surface
is small, and therefore the carrier density is low in the FeSC,
this does not mean that the Fermi level density of states, $N(E_F)$ is
low. In fact, it is high, and in particular higher than in cuprates.
Depending on the specific compound and details of the structure
used in the calculation, $N(E_F)$ can exceed 2 eV$^{-1}$ on a per Fe
both spins basis.
For example, the calculated value for LaFeAsO is $N(E_F)$=2.6 eV$^{-1}$.
Considering that the DOS near $E_F$ is mainly from Fe, this places
the FeSC near itinerant magnetism, based on nearness to the Stoner
criterion. Thus based on the band structure, these materials should
be characterized as
low carrier density, high density of states metals, near itinerant
magnetism.
This is in contrast to cuprate superconductors, which for optimum
doping are high carrier density, moderate density of states materials,
and are far from itinerant magnetism.
Consequences of the heavy bands in the FeSC relative to cuprates are
short superconducting coherence lengths and high superfluid densities.

\section{Magnetism}

Besides the high values of $N(E_F)$ in the FeSC, which place them
in proximity to itinerant magnetism in general, the Fermi surface
itself is nested. In particular, the approximately compensating
cylindrical hole and electron sections match reasonably well if
translated by ($\pi$,$\pi$,$k_z$) for arbitrary $k_z$, and in addition they
have similar $d_{xz}$/$d_{yz}$ orbital character. This leads to a peak
in the Lindhard susceptibility, \cite{mazin-spm,ma-mag,kuroki-spm,ishibashi}
which with Stoner enhancement is sufficient
to result in a magnetic instability.
This is a spin density wave (SDW)
instability of the Fermi surface driven by electrons at and near $E_F$.
This provides an explanation for the experimentally observed SDW order
found in most of the undoped FeSC.
\cite{cruz,rotter-sdw}
This SDW shows an evolution of the order parameter with temperature
consistent with the expected behavior of an SDW,
and optical spectroscopy shows gapping of the Fermi surface
and reduction of the scattering rate consistent with a Fermi surface
driven SDW. \cite{hu-sdw}
This is consistent with transport data, which show a
reduction in both resistivity and Hall carrier concentration
in the ordered state. 
\cite{mcguire-trans}
Importantly, although most of the Fermi surface is gapped by the
SDW in the FeSC, there are remaining carriers, and these are
unambiguously metallic, as shown by the observation of quantum
oscillations in the SDW state. \cite{sebastian}
This is in contrast to cuprates, where the undoped materials
are antiferromagnetic Mott insulators.

Furthermore, we emphasize that these materials are not near a
Mott insulator in the normal sense. In particular, a wide
range of chemical space has been explored so far, including
substitution of the counter-ions, such as various rare earths
in the oxy-pnictides, different alkaline earths in the ThCr$_2$Si$_2$
structure materials, replacement of As by P, and also chalcogens,
(S,Se,Te), and also alloying on the Fe site with Co and Ni.
This implies exploration of substantial parameter space, but
nonetheless, the compounds are practically all metallic and in particular no
Mott insulator has been found as a function of doping, pressure or chemical
substitution.

The density functional electronic structure
also provides a qualitative understanding of the phase diagrams.
In particular, magnetic order arises in general when the
Lindhard function, $\chi_0({\bf q},\omega=0)$
at the ordering wavevector ${\bf q}$ exceeds a critical value so that
there is a divergence of
the Stoner enhanced susceptibility,
$\chi_{S}({\bf q})=
\chi_0({\bf q})[1-I({\bf q})\chi_0({\bf q})]^{-1}$. Here
the Stoner enhancement parameter, $I({\bf q})$ is a smooth function of 
{\bf q}, reflecting the {\bf q} dependence of the band character,
while $\chi_0$ can be a strongly ${\bf q}$ dependent function reflecting
the Fermi surface.
Besides the Stoner enhancement,
in certain materials near a magnetic critical
point, as in e.g. weak itinerant magnets,
such as Ni$_3$Al and Sr$_3$Ru$_2$O$_7$,
\cite{aguayo,grigera}
there can be a suppression due to spin-fluctuations. In this regard,
high $\chi_{S}({\bf q})$ at ${\bf q}$ away from the ordering wavevector
can lead to spin fluctuations that compete with the ordered state.
\cite{aguayo,shimizu,moriya-book,mazin-zrzn2}

In any case, this sharp peak due to nesting of cylinders is at the 2D
wavevector that separates the axes of the cylinders, i.e. ($\pi$,$\pi$)
in the FeSC,
and with no $k_z$ dependence for true cylinders.
Thus one obtains a commensurate SDW at the zone corner. This is different
from Cr metal, where nesting vector is determined by a spanning vector
on the Fermi surface and is not set by symmetry,
yielding an incommensurate SDW. \cite{fawcett}
With doping the relative sizes of the hole and electron Fermi surface
sections must change in order to satisfy Luttinger's theorem.
For example, with hole doping the electron sheets
would become smaller, while
the hole surfaces would expand and in fact this is what is observed
experimentally. \cite{liu-pe}
For nested cylinders with radii differing by small $\delta q$, the sharp peak
at ($\pi$,$\pi$) will be smeared out to yield a plateau with
diameter $2\delta q$, thus reducing the peak value of $\chi_0$, while
retaining a large area integrated $\chi_0$ in the region around the 
zone corner. This is as found in calculations of the Lindhard
function for the doped FeSC. \cite{mazin-spm}
Thus the doping induced size mismatch between the electron and hole
Fermi surfaces will lower the peak value of $\chi_0$. Since this is
what controls magnetic ordering, it is qualitatively expected that
doping will work against the SDW state.

While this is what is observed generally in the FeSC, Fe$_{1+x}$Te
is an exception.
That material, which is doped by excess Fe outside the Fe square lattice
layers, shows an incommensurate SDW at high doping ($x$=0.165),
which becomes commensurate as the doping level is reduced. \cite{bao}
Density functional calculations show that FeTe, while sharing the same
basic electronic structure features with the FeSC, has larger Fermi surfaces,
with a greater tendency towards magnetism, and additionally that the
excess Fe in the doped compound carries magnetic moments, which may further
stabilize ordered states. \cite{subedi-fese,zhang-fete}
Significantly, if magnetism persists to high doping, where there is
a large size
mismatch $\delta q$ between the electron and hole Fermi surface sections
then an incommensurate SDW
is expected. In that case (specifically when radius of the smaller cylinder
becomes half the radius of the larger cylinder), a dip in the center of the
($\pi$,$\pi$) centered plateau in $\chi_0$ results, so that the maximum
in $\chi_0$ is no longer at the zone corner explaining the incommensurate SDW,
and leading to the expectation of a doping dependent wavevector.

This type of itinerant mechanism is distinct from local moment magnetism,
where moment formation is a result of on-site interactions and magnetism
is a consequence of weaker inter-site interactions, e.g. superexchange.
Itinerant magnetism differs from local moment magnetism in that the
physics involves the Fermi surface and is therefore long range, and in
the itinerant case
longitudinal degrees of freedom can be important, while in local moment
magnets only transverse degrees of freedom are important at low energy.

This is not to say that the Hund's coupling, which arises from
on-site atomic-like interactions, is not important in itinerant magnets.
It is after all the origin of the Stoner enhancement and
without it (i.e. $I({\bf q})$=0)
there would be no magnetic instability at all.
This is important
because it means that electronic states away from the Fermi energy
are involved in the magnetism and
will be coupled to the SDW through the exchange
interaction, and furthermore this may be observable in spectroscopy.
In fact, for bcc Cr \cite{fawcett} changes in the optical spectrum are
seen extending up to $\sim$ 0.5 eV upon the onset of the SDW.

However, while density functional calculations provide a qualitative
description of the properties of the FeSC, there are significant
quantitative problems. These show that the physics of these materials
is far from simple.
Experimentally, normal state of the doped superconducting
FeSC is paramagnetic, while the SDW ground states have varying magnetic
moments, but these are generally small, e.g. $\sim$ 0.4 $\mu_B$ in
LaFeAsO. \cite{cruz}
Therefore, one may expect that structural parameters, such as bond lengths,
obtained within density functional calculations without including magnetism
will be in good agreement with experimental values. This is not the case.
For example, in the prototype, LaFeAsO, the refined As height above the
Fe plane is 1.31 \AA, measured by neutron diffraction at 4 K, i.e. in
the ordered SDW state, 1.32 \AA, measured at 175 K, i.e.
in the metallic state above the SDW ordering temperature,
and 1.32 \AA, at 10 K, measured for the F doped superconducting material.
\cite{cruz}
In contrast, LDA calculations without including magnetism yield
1.16 \AA, i.e. an underestimate of $\sim$ 0.15 \AA. \cite{singh-du,mazin-mag}
Similar large
underestimates of the ligand height
are found for the other FeSC compounds, including FeSe. \cite{subedi-fese}

The problem is apparently magnetic in origin.
The FeSC show a very strong interplay of structure and magnetism as
was demonstrated by detailed calculations.
\cite{yin,mazin-mag}
This leads to an anomalous pressure dependence of the structure
with a collapse in the c-axis lattice parameter of CaFe$_2$As$_2$,
for example. \cite{kreyssig,yildirim-2}
In particular, the magnetism is strongly coupled to the ligand
height and the stability of magnetic states is greatly enhanced
as the ligand height is increased.
This interplay between magnetism and structure (via hybridization)
was discussed in a local picture by Wu and co-workers. \cite{wu-mag}
In any case, larger equilibrium ligand heights are obtained when magnetism is
included in the calculations, and quite reasonable agreement with
experiment is obtained, especially if calculations are done
using generalized gradient approximations (GGA), which more
strongly stabilize magnetism compared to the LDA for these compounds.
However, this comes at a price.
In particular, while the structure is improved, and the SDW type
ordering remains the predicted ground state the magnetism is
far too stable, with moments of $\sim$ 2 $\mu_B$ as compared
to experimental moments of less that 1 $\mu_B$ in most
compounds.
For comparison, standard GGA calculations for Fe metal give
a moment that is within 0.1 $\mu_B$ of experiment. \cite{singh-fe}
Furthermore, in the FeSC
a magnetic state is then incorrectly predicted to be the ground state
independent of doping.

Thus within density functional theory, the predicted ground state
of the FeSC is strongly magnetic, whereas experimentally the ground
state is a paramagnetic metal (doped) or a much
more weakly magnetic SDW state that can be destroyed in favor of
a paramagnetic superconductor with moderate pressure.
\cite{okada,fukazawa,alireza}
Furthermore, at least in LaFeAsO and other oxy-pnictides, the SDW is
preceded by a structural distortion that lowers the symmetry to that
of the SDW state, but at a higher temperature than the long range
magnetic orderings, \cite{cruz}
and importantly the largest signatures in transport and other properties
are at the structural transition.
The structural transition is explainable within DFT as a consequence
of the magnetic ordering (with large moments), \cite{yildirim,sushko}
and it has been
argued that its occurrence reflects a fluctuating magnetic state closely
related to the SDW but without long range static order.
\cite{mazin-j}

The type of error found in density functional calculations for the FeSC
is very different in nature from cuprates and other materials where
localization due to Coulomb repulsions plays the key role. In those
materials, the mean-field-like LDA treatment yields insufficient
localization and less tendency towards moment formation. This is the
case in the cuprates and in Mott insulators in general.
Cases where the LDA overestimates the tendency towards magnetism are
much less common and generally occur in materials near quantum critical
points, e.g. Ni$_3$Al and Sr$_3$Ru$_2$O$_7$.
These are materials where magnetic ordering is suppressed by quantum
fluctuations, which are beyond the scope of standard density functional
theory. Therefore we take the discrepancy of
ligand position as obtained in non-magnetic density functional theory from
the experimental value, in combination with the overly magnetic ground
state as evidence for the presence of strong spin fluctuations associated
with the SDW.
It is worth noting that at the GGA level, using experimental ligand positions,
these materials are unstable against magnetism in general, including
various antiferromagnetic configurations other than the SDW ordering,
though the SDW remains the lowest energy state. This may be important
in understanding the strong suppression of magnetism relative to
such calculations, since this general magnetic
instability does provide a large phase space for competing
fluctuations.

This is supported by several pieces of experimental evidence.
In several compounds, including both undoped and superconducing samples,
the susceptibility $\chi(T)$ is an increasing function of temperature
up to high temperatures.
\cite{mcguire-trans,klingeler-c,wang-c}
This shows the presence local antiferromagnetic correlations that persist
up to high temperature.
Fe exchange multiplets, demonstrating strong local fluctuating magnetism,
were found in non-magnetic superconducting CeFeAsO$_{0.89}$F$_{0.11}$
by x-ray absorption spectroscopy.
\cite{bondino}
The other case where this is observed in a non-magnetic Fe compound is
NbFe$_2$, \cite{vanacker}
which is an itinerant magnetic in close proximity to a 
quantum critical point. \cite{brando}
Furthermore, the scattering rate indicated by optics and by
transport is high in the normal state and decreases strongly
at the SDW onset. \cite{hu-sdw,mcguire-trans}

Returning to the Hund's coupling, the implication of the large dynamical
spin fluctuations that would needed to suppress the overly stable
magnetism in the mean-field-like GGA ground state, is that correspondingly
strong electronic signatures should be present.
This is consistent with the observation of exchange splitting of
$\sim$ 3 eV in the Fe 3$s$ spectra of paramagnetic CeFeAsO$_{0.89}$F$_{0.11}$.
\cite{bondino}
Considering that these are dynamical fluctuations, one may then
expect shifts in spectral weight
accompanied by increases in the scattering rate within the Fe derived
valence bands up to eV energies.

\section{Superconductivity}

At this time the mechanism for superconductivity has not yet been
established.
Detailed calculations have shown that the superconductivity
cannot be understood in terms of standard electron-phonon theory.
\cite{boeri,mazin-spm}
Also, there is a clear association, e.g. in the 
phase diagrams, between superconductivity and suppression of the SDW,
either by doping or by pressure.
Considering the evidence for spin fluctuations in the normal state
and the similarity of the normal states above the SDW transition and
above the superconducting transition it is tempting to consider
pairing based on spin fluctuations.
Key points are that the pairing interaction will have a shape closely
related to the real part of the susceptibility (but more strongly
peaked in weak coupling), and
that spin fluctuations are repulsive in a singlet channel. \cite{fay,allen}
Therefore the strongest interaction will be at ($\pi$,$\pi$) similar to
the SDW, and will favor opposite sign order parameters on Fermi surface
sections separated by this wavevector. Furthermore, the nesting
related peak in $\chi({\bf q})$ does not show strong
fine structure on the scale of the
small (electron or hole) Fermi surfaces
nor is there strong $k_z$ dependence. Therefore the itinerant
spin fluctuations do not provide a
driving force for changes in the order parameter within a given Fermi surface
section, either along $k_z$ or in the ($k_x$,$k_y$) plane.
Therefore, within the simplest scenario an $s$ symmetry state with opposite
sign order parameters on the electron and hole sections might be
expected. This is the $s_{\pm}$ state proposed
by Mazin and co-workers and by Kuroki and co-workers.
\cite{mazin-spm,kuroki-spm}
Experimental tests to determine the symmetry of the order parameter
are needed to confirm whether this is the actual superconducting state.
In any case, within such a framework, the same electrons
that drive the spin-fluctuations (i.e. those on the nested Fermi surface)
are the electrons that are involved in either the SDW order or the 
superconductivity. Therefore, both the SDW and superconductivity
are Fermi surface instabilities, driven by the same interaction and
competing for the same electrons.

As mentioned, both the superconducting and SDW orders arise
from a normal state that already has strong antiferromagnetic correlations
and so perhaps may be regarded as related. In this case, while there
is clearly a competition for electrons between the two orders, one
may ask whether they can co-exist.
A trivial possibility is that there is a first order line separating the
SDW and superconducting states, so that there is some region of two phase
co-existence, i.e. nanoscale phase separation with superconducting
and magnetic nanoregions. Even without a first order line,
local variations in the doping level may be expected to lead to a
mixed state with superconducting and magnetic regions on a nanoscale
near the boundary between these two states in the FeSC.
This is especially likely in the FeSC because of the heavy band
masses and resulting short coherence lengths, which mean that statistical
variations in dopant concentration within a volume defined by the
coherence length will be substantial. For example the estimated
in-plane coherence length of BaFe$_{1.8}$Co$_{0.2}$As$_2$ is
$\sim$ 28 \AA. \cite{yin-stm}
If the coherence length out-of-plane is half the in-plane value,
then the corresponding volume would be $\sim$ 10$^4$ \AA$^3$,
which would contain $\sim$ 200 transition metal atoms, and on average
20 Co atoms and therefore with local
$\sqrt{n}$ variations in the doping level of $\sim$ 20\%.

A more interesting possibility is that there may be a coexistence on
the Fermi surface.
In general, very soft fluctuations (energy below the gap) are pair
breaking in superconductivity since the superconducting state is unstable
against condensation of the fluctuations (i.e. magnetic order).
This could lead to a situation where part of the Fermi surface is gapped
by the condensation of spin fluctuations (i.e. magnetism) and
part is gapped by superconductivity, with a node separating these
regions.
This is somewhat analogous to one picture of the cuprate pseudogap state
based on angle resolved photoemission experiments. \cite{kondo-cu}
Such a state would be one way to reconcile conflicts between
experiments suggesting line nodes and experiments suggesting $s$
symmetry.
In this regard, we note that there is conflicting experimental evidence
regarding the gap in the FeSC.
Various experiments, including Andreev spectroscopy and photoemission
strongly suggest an $s$ wave gap,
\cite{ding-pe,wray-pe,kondo-pe,zhao-pe,chen-gap}
while at the same time nuclear magnetic resonance (NMR) measurements show
a $T^3$ dependence of the relaxation rate going to very low $T$. This
excess NMR relaxation rate
is generally a clear signature of line nodes, inconsistent with
a simple $s$ (including $s_{\pm}$) gap.
\cite{grafe-gap,matano-gap,nakai-gap,makuda,ahilan}
A similar conclusion is reached from penetration depth measurements,
i.e. that a non-exponential density of states is present.
\cite{gordon}
While an explanation in terms of a very strongly anisotropic $s$ wave
gap \cite{nagai-anis} may be possible, the persistence of the excess
NMR relaxation to very low temperature in samples with scattering
as indicated by the resistivity, may require a different explanation,
one possibility being the scenario outlined above.
Also, it should be mentioned that some of the data could perhaps be
explained by scattering effects. \cite{gordon}
In any case, it will be of great interest to examine the phase 
boundary between the superconducting and SDW states in detail.

\section{Summary and Conclusions}

The discovery of the FeSC provides us with the
second example of superconductivity above 50 K.
Thus the relationship between the FeSC and the cuprates is of considerable
interest. It is perhaps premature to make definitive connections
between these materials, considering that the pairing mechanism has
not been conclusively established in either material, and
in fact at this time not even the
pairing symmetry is clearly established in the FeSC.
Nonetheless, it still
of interest to note similarities and differences.

Starting with
similarities, neither the cuprates nor the FeSC can be understood
as conventional electron phonon superconductors and both materials have
both antiferromagnetic and superconducting phases in doping dependent
phase diagrams.
However, the differences appear to be more significant. In the cuprates
the antiferromagnetic phases are Mott insulators arising from the effects
of Coulomb correlations. These have substantial gaps and
local moment magnetism. In the FeSC the antiferromagnetic phases
are more directly connected with the normal state, as they are
metallic and arise from an SDW instability of the Fermi surface.
Unlike cuprates, superconductivity can be produced in the FeSC
by destroying the
SDW state without doping. Interestingly, the signatures of spin fluctuations
are much more apparent in the FeSC than in the normal state of the
cuprates, especially if one considers optimal doping.
Finally, much of the physics of the FeSC appears to be related to
itinerant electrons and the Fermi surface.

\section{Acknowledgements}

We are grateful for helpful discussions with I.I. Mazin, M.H. Du, H. Aoki,
Alaska Subedi, Lijun Zhang, A.S. Sefat, D. Basov,
D. Mandrus and B.C. Sales.
This work was supported by the Department of Energy,
Division of Materials Sciences and Engineering.

\bibliography{ferev}

\end{document}